\documentclass[10pt]{article}
\usepackage{graphicx}
\usepackage{amssymb} 
\usepackage[colorlinks,citecolor=purple,linkcolor=cyan,urlcolor=magenta]{hyperref}
\usepackage{xurl}
\usepackage[usenames,dvipsnames]{xcolor}
\usepackage{xspace}
\usepackage{lineno}
\usepackage{amsmath}
\usepackage{bm}
\usepackage{upgreek}
\usepackage[normalem]{ulem}
\usepackage[font=footnotesize,labelfont=bf]{caption}

\usepackage[super, comma]{natbib}
\newcommand*{\doi}[1]{\href{http://doi.org/#1}{#1}}
\makeatletter 
\renewcommand\@biblabel[1]{#1.} 
\makeatother

\usepackage{bibentry}

\usepackage{titlesec}
\titleformat*{\section}{\large\bfseries\sffamily}
\titlespacing\section{0pt}{12pt plus 4pt minus 2pt}{5pt plus 2pt minus 2pt}
\makeatletter
\def\ttl@useclass#1#2{%
  \@ifstar
    {\ttl@labelfalse\@dblarg{#1{#2}}}
    {\ttl@labeltrue\@dblarg{#1{#2}}}}
\makeatother

\usepackage{float}
\newfloat{extfig}{htbp}{} 
\floatname{extfig}{Extended Data Figure} 

\usepackage[a4paper, total={7.in, 10.2in}]{geometry}
\def\fg{$\mathsf{g}$\xspace}
\def\fb{$\mathsf{b}$\xspace}
\def\fr{$\mathsf{r}$\xspace}

\def\3c{3C\,279\xspace}
\newcommand{\ra}{\emph{RadioAstron}\xspace}
\newcommand{\aips}{\texttt{AIPS}\xspace}
\newcommand{\parsel}{\texttt{ParselTongue}\xspace}
\newcommand{\ehtim}{\texttt{eht-imaging}\xspace}
\newcommand{\difmap}{\texttt{DIFMAP}\xspace}
\newcommand{\smili}{\texttt{SMILI}\xspace}

\begin{document}

\nobibliography{references}

\noindent{\LARGE\bf\textsf{Filamentary structures as the origin of blazar jet radio variability}}

\bigskip\noindent
{Antonio Fuentes$^{1}$}, {José L. G\'omez$^{1}$}, {José M. Martí$^{2,3}$}, {Manel Perucho$^{2,3}$}, {Guang-Yao Zhao$^{1}$}, {Rocco Lico$^{1,4}$}, {Andrei P. Lobanov$^{5,6}$}, {Gabriele Bruni$^{7}$}, {Yuri Y. Kovalev$^{8,6,5}$}, {Andrew Chael$^{9,10}$}, {Kazunori Akiyama$^{11,12,13}$}, {Katherine L. Bouman$^{14}$}, {He Sun$^{14}$}, {Ilje Cho$^{1}$}, {Efthalia Traianou$^{1}$}, {Teresa Toscano$^{1}$}, {Rohan Dahale$^{1,15}$}, {Marianna Foschi$^{1}$}, {Leonid I. Gurvits$^{16,17}$}, {Svetlana Jorstad$^{18,19}$}, {Jae-Young Kim$^{20,21,5}$}, {Alan P. Marscher$^{18}$}, {Yosuke Mizuno$^{22,23,24}$}, {Eduardo Ros$^{5}$}, {Tuomas Savolainen$^{5,25,26}$}

\vspace{0.1cm}
\noindent
{\scriptsize
$^{1}$Instituto de Astrofísica de Andalucía (CSIC), Glorieta de la Astronomía s/n, 18008 Granada, Spain \\
$^{2}$Departament d'Astronomia i Astrofísica, Universitat de València, C/ Dr. Moliner, 50, E-46100 Burjassot, València, Spain \\
$^{3}$Observatori Astronòmic, Universitat de València, C/ Catedràtic José Beltrán 2, E-46980 Paterna, València, Spain \\
$^{4}$INAF—Istituto di Radioastronomia, via Gobetti 101, I-40129 Bologna, Italy \\
$^{5}$Max-Planck-Institut f\"ur Radioastronomie, Auf dem H\"ugel 69, D-53121 Bonn, Germany \\
$^{6}$Moscow Institute of Physics and Technology, Institutsky per. 9, Dolgoprudny, Moscow region, 141700, Russia \\
$^{7}$INAF—Istituto di Astrofisica e Planetologia Spaziali, via Fosso del Cavaliere 100, I-00133 Roma, Italy \\
$^{8}$Lebedev Physical Institute of the Russian Academy of Sciences, Leninsky prospekt 53, 119991 Moscow, Russia \\
$^{9}$Princeton Gravity Initiative, Princeton University, Jadwin Hall, Princeton, NJ 08544, USA \\
$^{10}$NASA Hubble Fellowship Program, Einstein Fellow \\
$^{11}$Massachusetts Institute of Technology Haystack Observatory, 99 Millstone Road, Westford, MA 01886, USA \\
$^{12}$National Astronomical Observatory of Japan, 2-21-1 Osawa, Mitaka, Tokyo 181-8588, Japan \\
$^{13}$Black Hole Initiative at Harvard University, 20 Garden Street, Cambridge, MA 02138, USA \\
$^{14}$California Institute of Technology, 1200 East California Boulevard, Pasadena, CA 91125, USA \\
$^{15}$Indian Institute of Science Education and Research Kolkata, Mohanpur, Nadia, West Bengal 741246, India \\
$^{16}$Joint Institute for VLBI ERIC (JIVE), Oude Hoogeveensedijk 4, 7991 PD Dwingeloo, The Netherlands \\
$^{17}$Aerospace Faculty, Delft University of Technology, Kluyverweg 1, 2629 HS Delft, The Netherlands \\
$^{18}$Institute for Astrophysical Research, Boston University, 725 Commonwealth Avenue, Boston, MA 02215, USA \\
$^{19}$Astronomical Institute, St. Petersburg State University, Universitetskij, Pr. 28, Petrodvorets, St. Petersburg 198504, Russia \\
$^{20}$Department of Astronomy and Atmospheric Sciences, Kyungpook National University, Daegu 702-701, Republic of Korea \\
$^{21}$Korea Astronomy and Space Science Institute, 776 Daedeok-daero, Yuseong-gu, Daejeon 34055, Republic of Korea \\
$^{22}$Tsung-Dao Lee Institute, Shanghai Jiao Tong University, Shengrong Road 520, Shanghai, 201210, People’s Republic of China \\
$^{23}$School of Physics and Astronomy, Shanghai Jiao Tong University, 800 Dongchuan Road, Shanghai, 200240, People’s Republic of China \\
$^{24}$Institut f\"ur Theoretische Physik, Goethe-Universit \"at Frankfurt, Max-von-Laue-Straße 1, D-60438 Frankfurt am Main, Germany \\
$^{25}$Aalto University Mets\"ahovi Radio Observatory, Mets\"ahovintie 114, FI-02540 Kylm\"al\"a, Finland \\
$^{26}$Aalto University Department of Electronics and Nanoengineering, PL15500, FI-00076 Aalto, Finland
}

\vspace{0.5cm}

\noindent
{\bf Supermassive black holes at the centre of active galactic nuclei power some of the most luminous objects in the Universe~\cite{Zensus1997}. Typically, very long baseline interferometric (VLBI) observations of blazars have revealed only funnel-like morphologies with little information of the ejected plasma internal structure~\cite{Jorstad2005, Lister2009a}, or lacked the sufficient dynamic range to reconstruct the extended jet emission~\cite{Kim2020}. Here we show microarcsecond-scale angular resolution images of the blazar \3c obtained at 22\,GHz with the space VLBI mission \ra~\cite{Kardashev2013}, which allowed us to resolve the jet transversely and reveal several filaments produced by plasma instabilities in a kinetically dominated flow. Our high angular resolution and dynamic range image suggests that emission features traveling down the jet may manifest as a result of differential Doppler-boosting within the filaments, as opposed to the standard shock-in-jet model~\cite{Marscher1985} invoked to explain blazar jet radio variability. Moreover, we infer that the filaments in \3c are possibly threaded by a helical magnetic field rotating clockwise, as seen in the direction of the flow motion, with an intrinsic helix pitch angle of $\sim$45$^{\circ}$ in a jet with a Lorentz factor of $\sim$13 at the time of observation.
}

\vspace{0.5cm}

\begin{figure*}[t!]
\centering
\includegraphics[width=1.\textwidth ]{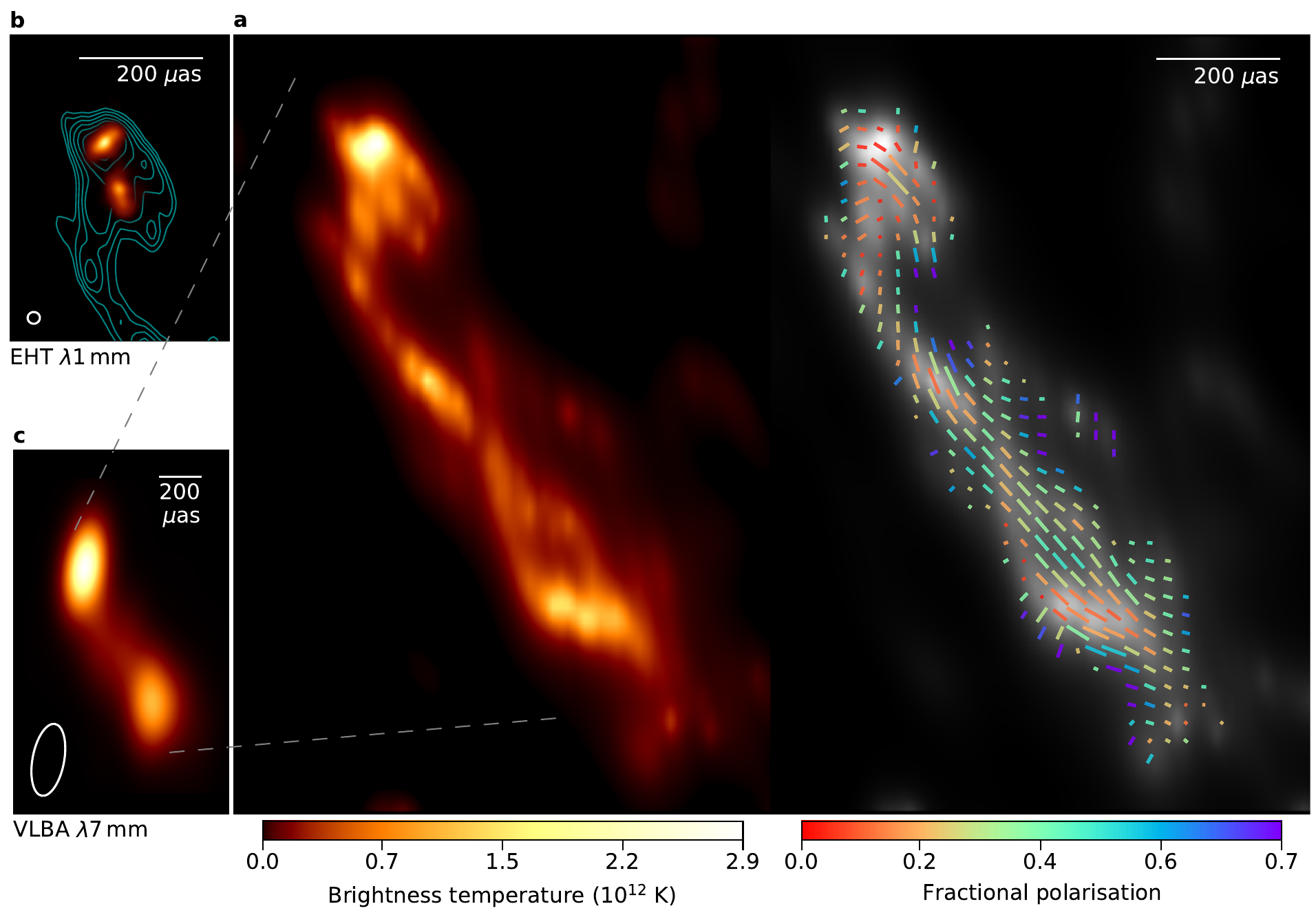}
\caption{{\bf The filamentary structure of the jet in \3c revealed by \ra. a}, Total intensity (left) and linearly polarized (right) \ra image at 1.3\,cm obtained on 10 March 2014. While both left and right images in \textbf{a} show brightness temperature in colour scale, the image on the right shows as well the recovered electric vector position angle overplotted as ticks. Their length and colour are proportional to the level of linearly polarized intensity and fractional polarization, respectively. {\bf b}, The 1:1 scale 1.3\,mm EHT image obtained in April 2017. Contours correspond to our \ra image, which are shown to compare the different scales probed. These start at 90\% of the peak brightness and decrease by successive factors of 3/2 until they reach 5\%. Both images were aligned with respect to the pixel with maximum brightness. {\bf c}, The 7\,mm VLBA-BU-BLAZAR program image obtained on 25 February 2014. White ellipses at the bottom-left corner of {\bf b} and {\bf c} indicate the $20\times20\,\upmu$as and $150\times360\,\upmu$as convolving beams, respectively. Bottom colour bars refer only to information displayed on {\bf a}.}
\label{fig:main}
\end{figure*}

We observed \3c on 10 March 2014 at 22\,GHz (1.3\,cm) with the space very long baseline interferometry (VLBI) mission \ra~\cite{Kardashev2013}, a 10-m space radio telescope (SRT) onboard of the \textit{Spektr-R} satellite, and an array of 23 ground-based radio telescopes spanning baseline distances from hundreds of kilometers to the Earth diameter (see \nameref{sec:methods} for a description of the array). The highly eccentric orbit of the SRT, with an apogee of $\sim350\,000$\,km, provided us with ground-space fringe detections of the source up to a projected baseline distance of 8 Earth diameters, probing a wide range of spatial frequencies perpendicular to the jet propagation direction (see Extended Data Fig.~\ref{fig:uvplot}). At the longest projected baselines to \ra, we achieved a resolving power of 27 microarcseconds ($\upmu$as), similar to that obtained by the Event Horizon Telescope (EHT) at 1.3\,mm ($\sim20\,\upmu$as)~\cite{Kim2020}. The large number of detections reported within the \ra ~ active galactic nuclei (AGN) survey program~\cite{Kovalev2020} made \3c an ideal target for detailed imaging. Fig.~\ref{fig:main} presents our \ra space VLBI polarimetric image of the blazar \3c. A representative image reconstruction obtained using novel regularized maximum likelihood methods~\cite{Chael2016, Chael2018} is shown along with the closest in time 7\,mm VLBA-BU-BLAZAR program image obtained on 25 February 2014, and the 1.3\,mm EHT image obtained in April 2017. We show a field of view of around $1\times1$ milliarcseconds (mas) with an image total flux density of $27.16\,$Jy, and note that all extended emission outside this region is resolved out by \ra. The robustness of our image is demonstrated in Extended Data Fig.~\ref{fig:fitting}, where we show how it fits the data used for both total intensity and linearly polarized image reconstruction. We acknowledge, however, that VLBI imaging is an ill-posed problem, and any image reconstruction that fits the data is not unique (e.g., see the comprehensive image analysis carried out in ref.~\cite{EHT4}). The image in Fig.~\ref{fig:main} is complemented by the 48 images presented in Extended Data Fig.~\ref{fig:survey} (see \nameref{sec:methods}).

\begin{figure}[t!]
\centering
\includegraphics[width=0.55\textwidth]{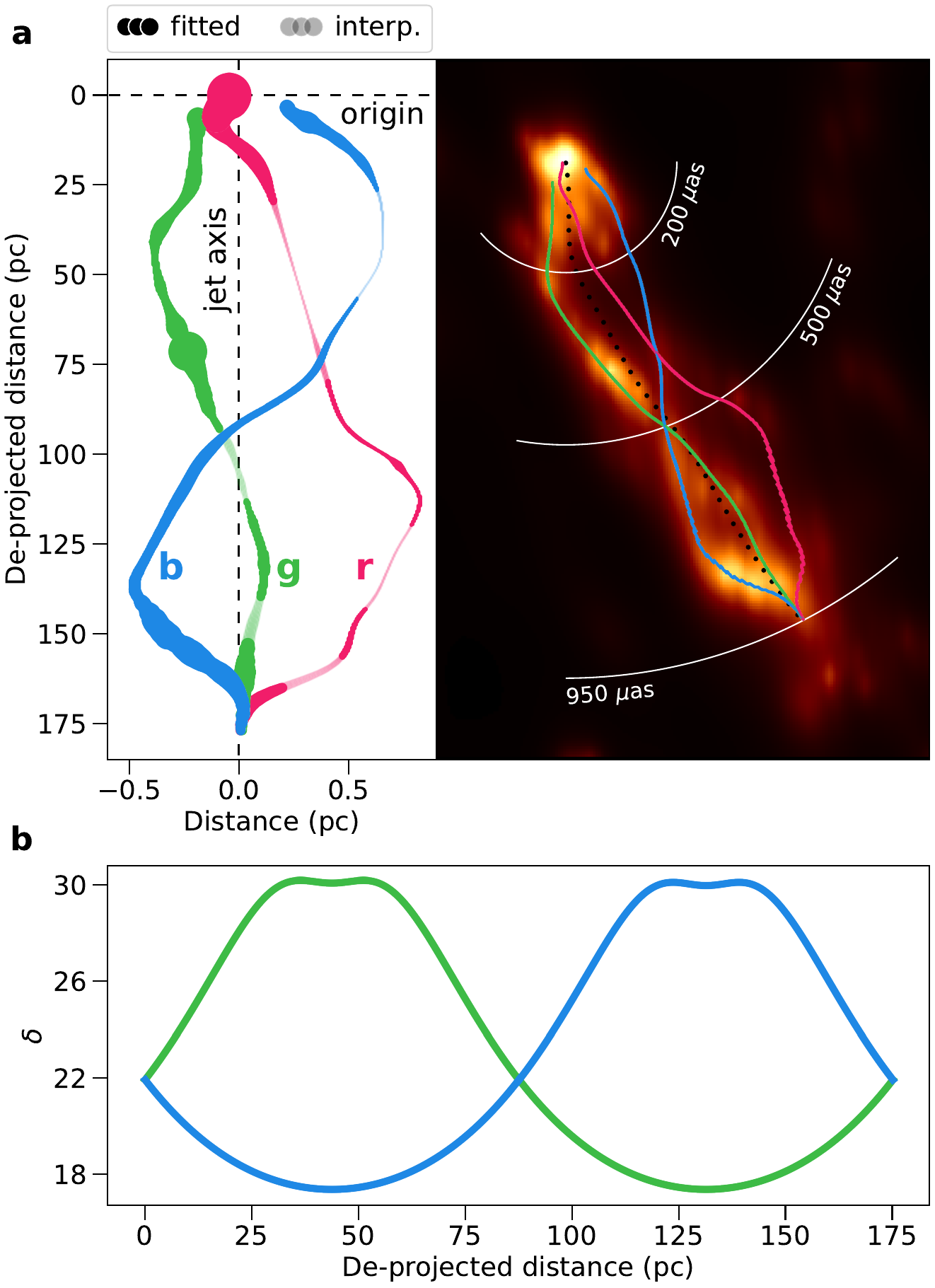}
\caption{\textbf{Analysis of the recovered filamentary structure. a}, Left: de-projected filament coordinates fitted using three Gaussian curves (see \nameref{sec:methods}). Dimmer points indicate regions where the coordinates have been interpolated. Marker size scales with the flux density. Note the scales in the x- and y-axis are different. Right: position of the fitted filaments on top of the reconstructed image. The black, dotted curve denotes the main jet axis.
\textbf{b}, Doppler factor computed for a plasma propagating along a double thread, as expected for elliptical modes, with $\Gamma=13$ and $\theta=1.9^\circ$.}
\label{fig:theory}
\end{figure}

In contrast to the contemporaneous 7\,mm and classical centimetre-wave VLBI jet images \cite{Lister2009a}, where the observed synchrotron emission seems to be contained in a funnel with a uniform cross section, we show in great detail the internal structure of a blazar jet and find strong evidence of the filamentary nature of the emitting regions within it. We identify the jet core as the upstream bright component, and the so-called `core region' encompasses the inner $\sim200\,\upmu$as, roughly the extent of the features probed at 1.3\,mm. The base is slightly elongated and tilted in the southeast-northwest direction, as reported in ref.~\cite{Kim2020}. However, contrary to the EHT sparse sampling of the Fourier plane \cite{Kim2020, EHT2}, the ground array supporting our space-VLBI observations provided a significantly larger filling fraction, which enabled us to reconstruct images with a dynamic range that is two orders of magnitude larger. Thus, while we can recover up to three different filaments emanating perpendicularly from the jet base, the EHT could only recover one and is ``blind" to the extended, filamentary structure, primarily due to the lack of short baselines. If aligned with respect to the brightness peak, both images match remarkably well, and the jet feature observed at 1.3\,mm is coincident in position and extension with our central filament, ignoring the small ($\lesssim 35\,\upmu$as) core shift between the two frequencies~\cite{Pushkarev2012}. Within our uncertainty, we do not measure a significant change in the core position angle with respect to the EHT image, taken three years later. The single-epoch results presented here do not allow us to discern whether this elongated structure corresponds to the accretion disk or to another extended jet component. Nonetheless, based on the small viewing angle inferred~\cite{Jorstad2017} ($\theta\sim1.9^\circ$) and the multi-epoch kinematic analysis of the model-fitted jet components, ref.~\cite{Kim2020} raised the possibility for this structure to correspond to a highly bent part of the inner jet.

Moving beyond the core region, we show in the top panel of Fig.~\ref{fig:theory} the de-projected and on-sky coordinates of the two main (hereinafter \fg and \fb), and possibly third (\fr), filaments obtained from the fitting of three Gaussian curves to transverse cuts to the main jet axis (see \nameref{sec:methods}). Further downstream, filament \fg is continuously recovered and contains most of the eastern extended structure flux density. Initially propagating in the southern direction, it displays a sharp bend of $\sim45^\circ$ to the west, close to the core region boundary. Although not continuously, we are also able to reconstruct filament \fb beyond the inner $200\,\upmu$as in what seems to be a helical-like morphology. These two filaments converge at $\sim500\,\upmu$as down the jet, where filament \fb crosses over \fg. Further downstream, they bend and converge again at $\sim950\,\upmu$as, where the brightness of the weaker filament is largely enhanced as it bends, dominating now the reconstructed emission in the southernmost jet region. Some diffuse emission is also systematically recovered parallel to filament \fg after the first crossing, which might indicate the presence of a third filament (\fr).

As detailed in the \nameref{sec:methods} section, there is a physically consistent domain in the space of parameters ($\Gamma$, $a_{\rm j}/a_{\rm ex}$), with $\Gamma$ within the previously determined range of jet flow Lorentz factor for \3c~\cite{Bloom2013} ($\Gamma\in[10,40]$), which allows us to interpret the observed approximate spatial periodicity, $\lambda_{\rm m}$, as the wavelength of the elliptical surface mode of a kinetically dominated, cold jet. More intriguing is the fact that the filaments associated with the elliptical mode are brighter in particular locations separated by half a wavelength ($\sim 400$ $\upmu$as for filament \fg, $\sim 850$ $\upmu$as for filament \fb), just before the crossings of the filaments. The properties of the flow (e.g., pressure, density, flow velocity) are modified locally by the elliptical wave, the magnitude of the changes depending on position and time as modulated by the wave phase. Such small changes in the properties of the flow could explain the differences in brightness between regions inside the jet and, in particular, along the filaments. Here, the perturbation in the three-velocity vector and the subsequent changes in the local Doppler boosting play a major role. The bottom panel of Fig.~\ref{fig:theory} shows how the Doppler factor, $\delta$, evolves along two threads originated by an elliptical mode in a plasma characterized by $\Gamma=13$ and $\theta=1.9^\circ$. With a ratio $\delta_{\rm max}/\delta_{\rm min}\simeq1.7$, the brightness of certain regions can increase by a factor of $\sim5$, in agreement with the brightness excess observed at $\sim 400$ $\upmu$as in filament \fg, and at $\sim 850$ $\upmu$as in filament \fb. This enhanced emission would then be the result of the Doppler boosted emission along the line of sight. This interpretation is supported by the fact that both enhanced emission regions are found at approximately the same phase of the corresponding helical filament, with the local flow velocity pointing along the same direction (the line of sight).

\begin{figure}[t!]
\centering
\includegraphics[width=0.6\columnwidth]{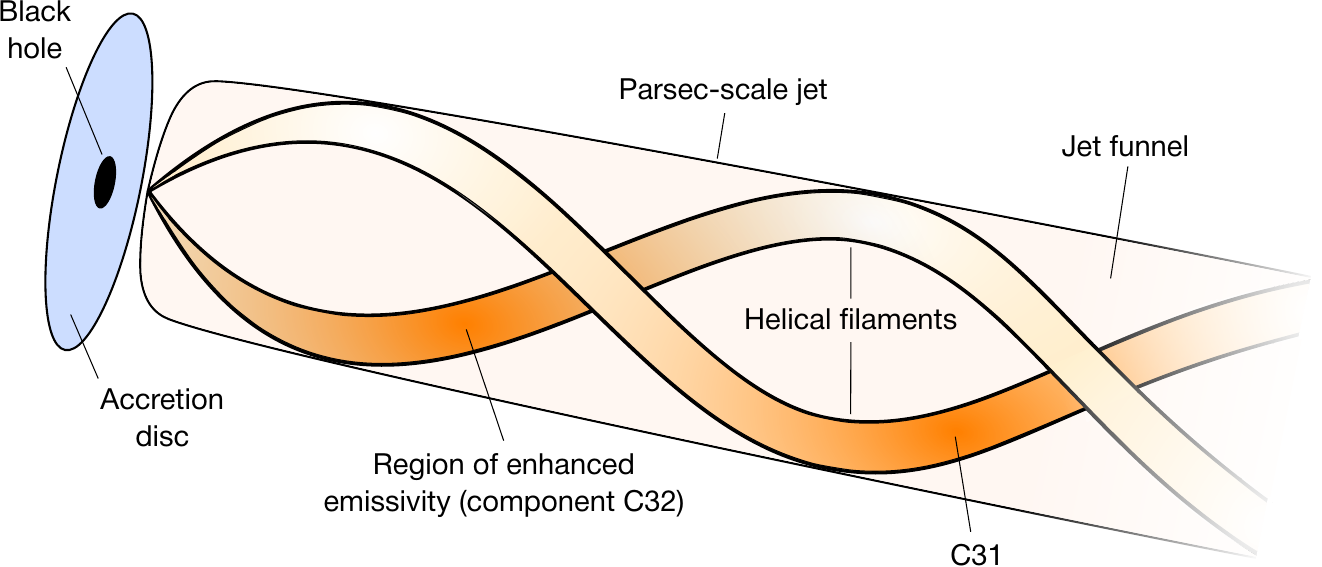}
\caption{Schematic figure of the proposed model for the internal jet structure in 3C\,279. The developing of plasma instabilities within the jet flow leads to filamentary structures with heterogeneous emissivity. Local changes in the plasma properties, which evolve and travel down the jet with the wave velocity, can explain these commonly observed moving components as a result of differential Doppler boosting.}
\label{fig:model}
\end{figure}

Continuing with this interpretation, it is important to note that the enhanced emission regions will not be steady but will propagate downstream at a (pattern) speed equal to the wave’s phase velocity. The fact that these brighter regions in filaments \fg and \fb match with the jet features observed at 7\,mm (components C32 and C31 in refs.~\cite{Jorstad2017,Weaver2022}, respectively) leads to the appealing possibility that these features correspond to the propagation of an elliptical perturbation mode and not to the propagation of a shock as proposed by the standard shock-in-jet model\cite{Marscher1985}. In particular, the estimated apparent speed of $\sim 7.23 \, c$ (component C31, ref.~\cite{Jorstad2017}), corresponding to a propagation speed of $\sim 0.996 \, c$ ($\Gamma\simeq11$) and close to the lower limit of the \3c jet Lorentz factor estimates, supports this possibility as these waves propagate downstream along the jet with velocities smaller than or equal to that of the jet bulk flow (see refs.~\cite{Hardee2000, Vega-Garcia2019} and \nameref{sec:methods}).
We illustrate our proposed model for the jet variability in \3c in the schematic figure shown in Fig.~\ref{fig:model}. Likewise, filamentary structures triggered by plasma instabilities can potentially explain the variability observed at radio wavelengths in other blazar sources and, in some cases, they could coexist with components originating from shock waves.
An important implication of our study is that the main emission occurs in thin filaments that cover only a fraction of the jet cross-section. The time-scale of variability of such features can therefore be much less than the light-travel time across the entire jet width, a fact that can help to explain extremely rapid variability~\cite{Hayashida2015}.
We note that, while our findings and proposed model come from the first spatially resolved image of a blazar internal structure, similar models based on differential Doppler boosting effects have been also suggested from indirect measurements (e.g., see ref.~\cite{Raiteri2017}). In the case of \3c, magnetic reconnection has been recently proposed~\cite{Shukla2020} as the mechanism behind gamma-ray flares, as opposed to shock waves or geometric effects. Nonetheless, the variability we model, as it originates from the formation of filamentary structures triggered in a kinetically dominated jet flow, concerns only that observed at radio wavelengths beyond the inner core region.

The analysis of the linearly polarized emission captured by \ra and the supporting ground array reveals clear signatures of a toroidal magnetic field threaded to the relativistic jet. The source is mildly polarized, with an integrated degree of linear polarization of $\sim$10\%. The electric vector position angle indicates a magnetic field predominantly perpendicular to the flow propagation direction, that is, consistent with a helical magnetic field dominated by its toroidal component. However, claiming the presence of a helical magnetic field usually implies finding rotation measure gradients across the jet width~\cite{Gabuzda2004,Broderick2010,Pasetto2021}. Although we find hints of transverse rotation measure gradients between our 22\,GHz observations and the closest in time VLBA-BU-BLAZAR 43\,GHz results~\cite{Weaver2022}, the large uncertainty associated with this preliminary analysis prevents us from drawing any robust conclusion. Relativistic magneto-hydrodynamic simulations of jets at parsec scales have shown that, in the presence of a helical magnetic field, the observed synchrotron emission is unevenly distributed across the jet width~\cite{Aloy2000, Fuentes2018, Moya-Torregrosa2021, Fuentes2021}. Based on the strong asymmetry in the reconstructed emission between the eastern and western sides of the jet axis, and following the analysis described in the \nameref{sec:methods} section, we can infer a jet bulk flow Lorentz factor of $\Gamma\simeq13$, which is in excellent agreement with the estimates provided by analyzing the kinematics of the parsec-scale jet. Moreover, this allow us to further infer a possible helical magnetic field, with an intrinsic pitch angle of $\sim45^\circ$, rotating clockwise as seen in the direction of flow motion.

The findings presented in this paper, supported as well by previous VSOP (e.g., ref.~\cite{Lobanov2001,Perucho2012}) and \ra (e.g., refs.~\cite{Bruni2021,Gomez2022}) space VLBI observations, strongly suggest that blazar jets have a complex and rich internal structure beyond the funnel-like morphologies reported by ground-based VLBI studies at lower angular resolutions. Future space VLBI missions and enhanced millimetre-wave global arrays, enabling high dynamic range observations capable to spatially resolving the jet width, should prove decisive in determining the true nature of jets powered by supermassive black holes.


\clearpage

\section*{Methods}
\label{sec:methods}

\noindent{\bf Observations}

\noindent Observations of \3c (1253$-$055) were conducted at 22.2\,GHz (1.3\,cm) on 2014 March 10-11, spanning a total of 11:44\,h from 14:15 to 01:59\,UT. During the observing session, \ra recorded evenly spaced (every 80$-$90\,min) blocks of data of 30\,min and one final block of $\sim$2\,h, corresponding to its orbit perigee. This allowed the spacecraft to cool down its high-gain antenna drive in between observing segments. Together with \ra, a ground array of 23 antennas observed the target, namely ATCA (AT), Ceduna (CD), Hobart (HO), Korean VLBI Network (KVN) antennas Tanman (KT), Ulsan (KU), and Yonsei (KY), Mopra (MP), Parkes (PA), Sheshan (SH), Badary (BD), Urumqi (UR), Hartebeesthoek (HH), Kalyazin (KL), Mets\"ahovi (MH), Noto (NT), Torun (TR), Medicina (MC), Onsala (ON), Yebes (YS), Jodrell Bank (JB), Effelsberg (EF), Svetloe (SV), and Zelenchukskaya (ZC).

Left and right circularly polarized signals (LCP and RCP, respectively) were recorded simultaneously at each station, with a total bandwith of 32\,MHz per polarization. Collected data were then processed at the Max-Planck-Institut f\"ur Radiostronomie using the upgraded version of the DiFX correlator~\cite{Bruni2016}. Fringes between \ra and ground stations were searched using the largest dishes, separately for each scan. This provides a first-order clock correction, to be later refined with baseline stacking in \aips~\cite{Greisen1990}. When no signal was found, we adopted a best-guess clock value extrapolated from scans giving fringes, with the aim of performing a further global fringe search at a later stage with \aips.

\bigskip\noindent{\bf Data reduction}

\noindent For the initial data reduction, we made use of \parsel~\cite{Kettenis2006}, a Python interface for \aips. At a first stage, we performed an \textit{a priori} calibration of the correlated visibility amplitudes using the system temperatures and gain curves registered at each station. Some of the antennas participating in the observations failed to deliver system temperature information, which we compensated by using nominal values modulated by the antenna's elevation at each scan. Since we chose the average system temperature as the station's default value, visibility amplitudes were not properly scaled. We overcame this issue by determining, every two minutes, the gain corrections needed for each IF and polarization from a preliminary image where only closure-quantities (closure phases and log-closure amplitudes) were involved, using the software library \smili~\cite{Akiyama2017a, Akiyama2017b}. Then, we applied to each antenna the mean gain value obtained, allowing for further residual corrections during the final imaging and self-calibration. The image total flux density was fixed to that measured by the intra-KVN baselines (27.65\,Jy), whose \textit{a priori} calibration was excellent~\cite{Cho2017}. Finally, we corrected the phase rotation introduced by the receiving systems as the source's parallactic angle changes.

We then solved for residual single- and multi-band delays, phases, and phase rates by incrementally fringe-fitting the data. In the first iteration we excluded \ra and performed a global fringe search on the ground array with a solution interval of 60\,s, using MP and EF as reference antennas for the first and second part of the experiment, respectively. Once fully calibrated, the ground array was coherently combined (trough baseline stacking) to increase the signal-to-noise ratio of possible fringe detections to \ra. To account for the acceleration of the spacecraft near its perigee and the low sensitivity of the longest projected baseline lengths to it, we adopted different solution intervals (from 10\,s to 240\,s) and data total bandwidth (by combining IFs).
With a signal-to-noise ratio cutoff of 5, reliable ground-space fringes were detected up to $\sim$8 Earth's diameters, corresponding to the first observing block of \ra (around 14\,UT), achieving a maximum angular resolution of 27\,$\upmu$as in the transverse direction to the jet axis. Lastly, we solved for the antennas' bandpass, the delay difference between polarizations using the task RLDLY, and exported the frequency averaged data along each IF. The fringe-fitted visibility coverage in the Fourier plane is shown in Extended Data Fig.~\ref{fig:uvplot}.

\bigskip\noindent{\bf Imaging}

\noindent Imaging of the data was carried out using novel regularized maximum likelihood (RML) methods~\cite{Narayan1986}, implemented in the \ehtim software library~\cite{Chael2016, Chael2018}.
While the CLEAN algorithm~\cite{Hogbom1974} has been widely used in the past for VLBI image reconstruction, novel RML methods are not extensively used, especially at centimetre wavelengths and space VLBI experiments. Generally speaking, RML methods try to solve for the image $I$ that minimizes the objective function:
\begin{linenomath*}
\begin{equation}\label{eq:obj_fun}
    J(I) = \sum_{\rm{data\ terms}}\alpha_D\chi^2_D(I, V) - \sum_{\rm{reg.\ terms}}\beta_R S_R(I),
\end{equation}
\end{linenomath*}
where $\alpha$ and $\beta$ are hyperparameters that weight the contribution of the image fitting to the data $\chi^2$, and the image-domain regularization $S$, to the minimization of the previous equation. Contrary to traditional CLEAN, full closure data products (closure phases and log closure amplitudes) can be employed during image reconstruction in addition to complex visibilities, further constraining the proposed image. Given the large number of telescopes participating in the experiment, closure quantities have proven quite useful since atmospheric phase corruption and gain uncertainties are mitigated. Multiple regularization over the proposed image can be imposed too, like smoothness between adjacent pixels or similarity to a prior image.

Prior to imaging, we first performed an initial phase-only self-calibration to a point source model with a solution interval of 5\,s and coherently averaged the data in 120\,s intervals, using the \difmap package~\cite{Shepherd1997}. We compared these results with those obtained with the \aips task CALIB, for which a signal-to-noise ratio cutoff of 5 was set, to ensure no artificial signal was introduced in the data. In the following paragraphs we describe the imaging procedure.

As a first a step, we flagged all baselines to \ra and imaged the data collected only by ground radio telescopes. The pre-processed data noise budget is inflated by a small amount (1.5\,\%), to account for non-closing errors, and the image is initialized with an elliptical Gaussian, oriented in roughly the same angle as the 7\,mm image and enclosed in a $1.5\times1.5$\,mas field of view gridded by 200$\times$200 pixels.
As mentioned above, because of the poor \textit{a priori} amplitude calibration due to missing antennas' system temperature, we opted for a first round of imaging where only closure quantities (closure phases and log closure amplitudes) were used to constrain the image likelihood. This likelihood takes the form of the mean squared standardized residual (similar to a reduced $\chi^2$) as defined in ref. \cite{EHT4}. Each imaging iteration takes as initial guess the image reconstructed in the previous step blurred to the ground array nominal resolution, i.e., 223 $\upmu$as, which prevents the algorithm of being caught in local minima during optimization of Equation~\ref{eq:obj_fun}. We then self-calibrate the data to the closure-only image obtained and incorporate full complex visibilities to the imaging process, which is finalized by repeating the imaging and self-calibration cycle two more times. In addition to the data products mentioned, we impose several regularizations to the proposed images. These include maximum entropy (mem), which favors similarity to a prior image; total variation (tv) and total squared variation (tv2), which favor smoothness between adjacent pixels; $\ell_1-$norm, which favors sparsity in the image; and total flux regularization, which encourages a certain total flux density in the image. Finally, we restore all baselines to \ra and repeat this procedure, substituting the Gaussian initialization with the blurred, ground-only image previously reconstructed and using the full array nominal resolution (27 $\upmu$as) between imaging iterations to blur intermediate reconstructions.

Contrary to full Bayesian methods, RML techniques do not estimate the posterior distribution of the underlying image, but instead compute the maximum a posteriori solution, i.e., the single image that best minimizes Equation~\ref{eq:obj_fun}. The hyperparameters chosen will necessarily have an impact on the reconstructed image features, thus we conducted a scripted parameter survey to ensure the robustness of the subtle structures seen in Fig.~\ref{fig:main} and to impartially determine which parameters perform better on the image reconstruction. From the many images obtained, we show in Extended Data Fig.~\ref{fig:survey} the complete collection of images which could potentially describe the observed source structure. These fit the data equally well and preserve the total flux measured by the KVN to a certain level. The regularizers and hyperparameters used to obtain these images are listed on each panel of Extended Data Fig.~\ref{fig:survey}. Apart from these, we gave the same weight to complex visibilities and closure quantities data terms. Although we can observe some differences in the weaker emission, the main filamentary structure is present in all the images. Fig.~\ref{fig:main} corresponds to \#21, which has the overall minimum reduced $\chi^2$.

\bigskip\noindent{\bf Synthetic data tests}

Given that one of the helical filaments reconstructed in our images (filament \fb) crosses over the other with a position angle roughly oriented in the direction where we lack space-ground baseline coverage, we tested the ability of the image reconstruction algorithm to successfully recover similarly oriented structures. With that aim, we generated two synthetic data sets using \ehtim (e.g., see refs.~\cite{EHT4,EHT_SGRA_III}). These simulate space VLBI observations of the source models presented in Extended Data Fig.~\ref{fig:syn_data} with the same baseline coverage as that of Extended Data Fig.~\ref{fig:uvplot}. The data sets are corrupted with thermal noise and visibility phases are scrambled. These simple geometric models are specially designed to mimic the crossing of two helical filaments and to some extent the brightness distribution of the \3c images reconstructed. While the first model is oriented as the parsec scale jet in \3c, the second model is oriented perpendicularly to the fitted beam position angle ($\sim10^\circ$).

In Extended Data Fig.~\ref{fig:syn_data} we show the images reconstructed following the procedure described in the previous section. For comparison, we also present the images obtained after flagging all \ra baselines, that is, with only the ground array. When the full array is considered, we are able to fully recover the crossing of the filaments and ground-truth structure of the model oriented as the jet in \3c, although we acknowledge a worse performance reconstructing the internal structure of the model oriented perpendicularly to the position angle of the fitted beam, just as expected. On the contrary, when space-ground baselines are removed, the images reconstructed display a jet correctly oriented but with no traces of helical filaments. Thus, we are confident that the recovered filamentary structure, only achievable at this observing frequency thanks to \ra, is derived from the intrinsic source structure and not from the lack of North-South space-ground baselines.

\bigskip\noindent{\bf Polarimetric imaging}

\noindent The polarization results presented in Fig.~\ref{fig:main} were obtained using the \ehtim library as well. A more complete description of the method can be found in refs. \cite{Chael2016, EHT7}, here we briefly outline the procedure followed. For polarimetric imaging, \ehtim minimizes again Equation~\ref{eq:obj_fun}, substituting complex visibilities and closure quantities data terms by polarimetric visibilities $\mathcal{P}=\mathcal{\tilde{Q}}+i\,\mathcal{\tilde{U}}$ and the visibility domain polarimetric ratio $\breve{m}=\mathcal{P}/\mathcal{\tilde{I}}$. Note that total intensity and linearly polarized intensity images are reconstructed independently. Image regularization includes now total variation which, as for total intensity imaging, encourages smoothness between adjacent pixels; and the Holdaway-Wardle regularizer \cite{Holdaway1990}, which prefers pixels with polarization fraction values below the theoretical maximum 0.75. The pipeline then alternates between minimizing the polarimetric objective function and solving for the complex instrumental polarization, the so-called D-terms. The instrumental polarization calibration is performed by maximizing the consistency between the self-calibrated data and sampled data from corrupted image reconstructions. After D-terms solutions are found for each antenna, the reconstructed polarimetric image is blurred, as was done for Stokes $\mathcal{I}$ imaging, and the imaging-calibration cycle is repeated until convergence of the solutions. 

Apart from instrumental polarization, VLBI polarimetric analyses rely on the calibration of the absolute polarization angle. To account for this, we compared our polarization results with the closest in time 7\,mm results. The recovered polarization angle patterns match remarkably well when our image is convolved with the 7\,mm beam. Ignoring Faraday rotation of the polarization angle between the two frequencies, based on the small rotation measure values reported in refs~\cite{Hovatta2012, Park2018}, we estimate an overall median difference of $\sim8^\circ$, that we applied to the results presented in Fig.~\ref{fig:main}.

\bigskip\noindent{\bf Filament fitting}

\noindent The relative right ascension and declination coordinates of the filaments were obtained from the fitting of three Gaussian curves to transverse profiles of the brightness distribution. We first computed the main jet axis, commonly referred to as ridge line, from a convolved version of the reconstructed image. Using a sufficiently large Gaussian kernel, we blurred our image until the emission blends into a unique stream and the filaments are no longer distinguishable, similarly to the 7\,mm VLBA-BU-BLAZAR image. We then project this image into polar coordinates, centred at the jet origin, and slice it horizontally, storing the position of the flux density peak for each cut. These positions are then transformed back to Cartesian coordinates, obtaining thus the main jet axis. To each pair of consecutive points conforming the axis, we compute the local perpendicular line and retrieve the flux density of the pixels contained in the cut. With this procedure, we assemble a set of transverse brightness profiles to which we fit the sum of three Gaussian curves using the python package \texttt{lmfit} \cite{lmfit}. The number of Gaussian components used is motivated by the number of filaments observed emanating from the core region, although we note that two Gaussian components are enough to fit the two main threads. Finally we select the coordinates as the position of the peak(s) found in the curve best fitting each cut. In Figure~\ref{fig:fitting}, coordinates are de-projected assuming a source redshift $z=0.536$ \cite{Marziani1996}, a viewing angle $\theta=1.9^\circ$ \cite{Jorstad2017}, and a cosmology $H_0=67.7$\,km\,s$^{-1}$\,Mpc$^{-1}$, $\Omega_m=0.307$, and $\Omega_\Lambda=0.693$ \cite{Planck2016}.

\bigskip\noindent{\bf Instability analysis}

\noindent Based on the aforementioned Gaussian fitting to the observed filaments, we estimate an approximate spatial periodicity $\lambda_{\rm m}$ of 950\,$\upmu$as (projected on the plane of the sky) or 175\,pc (de-projected), which corresponds to $\sim2.3\times 10^6$ gravitational radii assuming a black hole mass of $M_{\rm BH}\simeq8\times10^8\,M_{\odot}$ \cite{Nilsson2009}.
The possibility for these filaments to reflect a fundamental periodicity of the black hole or inner accretion disk directly associated with their rotation should be dismissed, as it would imply propagation speeds along the filaments larger than the speed of light by orders of magnitude. At the same time, explaining such a fundamental periodicity in terms of precession of a jet nozzle, caused by the Lense-Thirring effect \cite{Bardeen1975} or a supermassive black hole binary system, invoked to explain a sharp bend in the nuclear region, have been recently discarded~\cite{Kim2020}. On the other hand, anchoring the filaments to the outer accretion disk to allow for a subluminal propagation of the helical pattern would imply a exceedingly large (Keplerian) disk radius, that is, larger than $\sim$ 1 light-year, about two orders of magnitude larger than the expected disk sizes~\cite{Morgan2010}.

According to ref.~\cite{Kim2020}, the jet no longer accelerates beyond $\sim$100\,$\upmu$as from the core, suggesting a kinetically dominated flow in which the observed filaments show a magnetic field structure dominated by the toroidal component. Taking this into account, we conclude that these bright filaments reveal compressed regions with enhanced gas and magnetic pressure -- favouring an increased synchrotron emissivity and ordering of the magnetic field. Thus, these might be associated with the triggering and development of flow instabilities. Current-driven kink or Kelvin-Helmholtz (KH) instabilities are the most plausible mechanisms capable of developing such helical structures~\cite{Mizuno2012, Perucho2012, Vega-Garcia2019}. Rayleigh-Taylor (RT) instabilities have also been discussed in the context of jet expansion and recollimation, as a possible trigger of small-scale distortions of the jet surface and turbulent mixing (see ref.~\cite{Perucho2019} for a review). However, RT instabilities would not produce filaments as the ones we observe, so we neglect this option. Current-driven instabilities dominate in Poynting-flux regimes with strong helical magnetic fields, that is, in the jet's acceleration and collimation region. On the contrary, KH instabilities have the largest growth in kinetically dominated flows, thus favored in our case.
The extension of the filaments greatly exceeds the jet radius, which is expected for KH surface modes. While two filaments could be generated by an elliptical mode, the possible third filament observed might indicate the presence of an additional helical mode interfering with the elliptical.

Assuming the jet is kinetically dominated and cold, as expected for powerful jets already expanded and accelerated, the fastest growing frequency of a mode is given by $\omega^*_{nm}R/a_{\rm ex} = (n+2m+1/2)\pi/2$~\cite{Hardee1987, Hardee2000}, where $R$ is the radius, $a_{\rm ex}$ the sound speed of the ambient medium, and $n$ and $m$ the type of mode ($n=1, 2$ for helical and elliptical modes, respectively; and $m=0$ for a surface mode). Taking $\omega \leq 2\pi c/\lambda_{\rm m}$, we find $a_{\rm ex} \approx 10^{-2} c$ for both the helical and elliptical modes. At this maximum growth frequency, and for a highly supersonic jet (i.e., with jet Mach number $M_{\rm j}\gg1)$, the wavelength of the mode and wave velocity are given, respectively, by
\begin{linenomath*}
\begin{align}
\lambda_{\rm m}^*& \approx \frac{4}{n + 1/2}\,M_{\rm ex} \frac{\Gamma}{a_{\rm j}/a_{\rm ex} + \Gamma}\,R \quad\rm{and} \label{eq:wavf1}\\
v^*_{\rm w}&\approx\frac{\Gamma}{a_{\rm j}/a_{\rm ex} + \Gamma}\,u,  \label{eq:wavf2}
\end{align}
\end{linenomath*}
where $u$ is the jet flow velocity (which approximates the light speed $c$ given the large Lorentz factors inferred), $\Gamma$ is the jet flow Lorentz factor, $M_{\rm ex}$ is the Mach number of the jet with respect to the ambient sound speed ($M_{\rm ex} = u/ a_{\rm ex} \approx c/a_{\rm ex} \approx 100$), and $a_j$ is the sound speed of the jet flow.

In our interpretation, described in the main body of the paper, the wave velocity $v^*_{\rm w}$ coincides with the (pattern) speed of the jet feature observed at 7 mm (component C31 in ref.~\cite{Jorstad2017}), very close to $c$, leading to the condition (see Eq.~\ref{eq:wavf2}) $a_{\rm j}/a_{\rm ex} \ll \Gamma$ or, equivalently, $M_{\rm j} \gg M_{\rm ex}/\Gamma$. With $M_{\rm ex} \approx 100$ and $\Gamma \in [10,40]$, the last condition implies $M_{\rm j} \gg 1$, hence validating our assumption of a kinetically dominated flow at the observed scales.

\bigskip\noindent{\bf Jet properties derived from the reconstructed polarimetric emission}

\noindent The synchrotron radiation coefficients are a function of the angle between the magnetic field and the line of sight in the fluid frame. Thus, for a fixed viewing angle and jet flow velocity, the bulk of the emission will be located on either side of the main jet axis depending on the magnetic field helical pitch angle. This asymmetry is maximized when the helical magnetic field pitch angle (in the fluid frame) $\phi'$ equals to $45^\circ$ (ref.~\cite{Aloy2000}). Given the strong asymmetry in the reconstructed emission between the eastern and western sides of the jet axis, we can assume that the viewing angle in the fluid's frame approximates $\phi'$, that is $\theta'\simeq\phi'$. Hence, given the estimated viewing angle in the observer's frame ($\theta\sim1.9^\circ$) (ref.~\cite{Jorstad2017}) and the light aberration transformations~\cite{Rybicki1979}
\begin{linenomath*}
\begin{align}
\sin\theta' = \frac{\sin\theta}{\Gamma(1-\beta\cos\theta)}, \quad \cos\theta' = \frac{\cos\theta-\beta}{1-\beta\cos\theta}, \label{eq:light}
\end{align}
\end{linenomath*}
where $\beta=\sqrt{1-1/\Gamma^2}$, we can infer a jet bulk flow Lorentz factor of $\Gamma\simeq13$, which is in excellent agreement with the estimates provided by analyzing the kinematics of the parsec-scale jet~\cite{Jorstad2017}, and satisfies the upper limit previously established by our KH instability analysis. Moreover, this allows us to estimate the viewing angle $\theta_r$ at which the emission asymmetry will reverse from one side to the other as $\cos\theta_r = (1-1/\Gamma^2)^{1/2}$ (ref.~\cite{Aloy2000}), which results in $\theta_r\simeq4.4^\circ$ for $\Gamma=13$. Since $\theta<\theta_r$ and the bulk of the reconstructed emission is located to the east of the jet axis, we infer a helical magnetic field rotating clockwise as seen in the direction of flow motion. The Lorentz-transformation of the magnetic field from the fluid's to the observer's frame boosts the toroidal component by $\Gamma$ (ref.~\cite{Lyutikov2005}), and therefore the helix pitch angle transforms as $\tan\phi=\Gamma\tan\phi'$. This makes $\phi\simeq86^\circ$ in the observer frame, which is in agreement with the predominantly toroidal magnetic field observed.


\section*{Data availability}

\noindent Pre-processed data used for imaging is available at \url{https://github.com/aefez/radioastron-3c279-2014}.


\section*{Code availability}

\noindent The software packages used to calibrate, image, and analyze the data are available at the following sites: \aips (\url{http://www.aips.nrao.edu/index.shtml}), \parsel (\url{https://www.jive.eu/jivewiki/doku.php?id=parseltongue:parseltongue}), \difmap (\url{https://science.nrao.edu/facilities/vlba/docs/manuals/oss2013a/post-processing-software/difmap}), \smili (\url{https://github.com/astrosmili/smili}), \ehtim (\url{https://github.com/achael/eht-imaging}), and \texttt{lmfit} (\url{https://lmfit.github.io/lmfit-py/}).



\section*{Acknowledgements}

\noindent We thank L. Hermosa for useful comments on the manuscript. The work at the IAA-CSIC is supported in part by the Spanish Ministerio de Econom\'{\i}a y Competitividad (grants AYA2016-80889-P, PID2019-108995GB-C21), the Consejer\'{\i}a de Econom\'{\i}a, Conocimiento, Empresas y Universidad of the Junta de Andaluc\'{\i}a (grant P18-FR-1769), the Consejo Superior de Investigaciones Cient\'{\i}ficas (grant 2019AEP112), the State Agency for Research of the Spanish MCIU through the ``Center of Excellence Severo Ochoa" award to the Instituto de Astrof\'{\i}sica de Andaluc\'{\i}a (SEV-2017-0709), and the grant CEX2021-001131-S funded by MCIN/AEI/10.13039/501100011033.
JMM and MP acknowledge support from the Spanish \emph{Ministerio de Ciencia} through grant PID2019-107427GB-C33, and from the \emph{Generalitat Valenciana} through grant PROMETEU/2019/071. JMM acknowledges additional support from the Spanish \emph{Ministerio de Econom\'{\i}a y Competitividad} through grant PGC2018-095984-B-l00. MP acknowledges additional support from the Spanish \emph{Ministerio de Ciencia} through grant PID2019-105510GB-C31. MP acknowledges Manuel Perucho-Franco, his father, for standing up as an example through his whole life.
YYK was supported by the Russian Science Foundation grant 21-12-00241.
AC is supported by Hubble Fellowship grant HST-HF2-51431.001-A awarded by the Space Telescope Science Institute, which is operated by the Association of Universities for Research in Astronomy, Inc., for NASA, under contract NAS5-26555.
JYK was supported for this research by the National Research Foundation of Korea (NRF) grant funded by the Korean government (Ministry of Science and ICT; no. 2022R1C1C1005255).
YM acknowledges support from the National Natural Science Foundation of China (grant no. 12273022) and Shanghai pilot program of international scientist for basic research (grant no. 22JC1410600).
TS was supported by the Academy of Finland projects 274477, 284495, 312496, and 315721.
The \ra project is led by the Astro Space Center of the Lebedev Physical Institute of the Russian Academy of Sciences and the Lavochkin Scientific and Production Association under a contract with the Russian Federal Space Agency, in collaboration with partner organizations in Russia and other countries. 
The European VLBI Network is a joint facility of independent European, African, Asian, and North American radio astronomy institutes. Scientific results from data presented in this publication are derived from the EVN project code GA030D.
This research is partly based on observations with the 100\,m telescope of the MPIfR at Effelsberg. 
This publication makes use of data obtained at Mets\"ahovi Radio Observatory, operated by Aalto University in Finland.
Our special thanks to the people supporting the observations at the telescopes during the data collection.
This research is based on observations correlated at the Bonn Correlator, jointly operated by the Max-Planck-Institut f\"ur Radioastronomie, and the Federal Agency for Cartography and Geodesy. 
This study makes use of 43\,GHz VLBA data from the VLBA-BU Blazar Monitoring Program (VLBA-BU-BLAZAR; \url{http://www.bu.edu/blazars/BEAM-ME.html}), funded by NASA through the Fermi Guest Investigator grant 80NSSC20K1567. \\

\noindent The Version of Record of this article is published in {\it Nature Astronomy}, and is available online at \url{https://doi.org/10.1038/s41550-023-02105-7}.


\section*{Author contributions}

\noindent A.F., J.L.G., and G.Y.Z. worked on the data calibration. A.F., J.L.G., G.Y.Z., R.L., A.C., K.A., K.L.B, H.S., I.C, and E.T. worked on the image reconstruction and analysis. G.B. correlated the space VLBI data. J.M.M., M.P., A.F., J.L.G, and Y.M worked on the interpretation of the results. All authors contributed to the discussion of the results presented and commented on the manuscript. 


\section*{Competing interests}

\noindent The authors declare no competing interests.


\section*{Additional information}

\noindent Correspondence and requests for materials should be addressed to Antonio Fuentes (\href{mailto:afuentes@iaa.es}{afuentes@iaa.es}) or José L. Gómez (\href{mailto:jlgomez@iaa.es}{jlgomez@iaa.es})

\newpage

\begin{extfig*}
\centering
\includegraphics[width=0.7\textwidth]{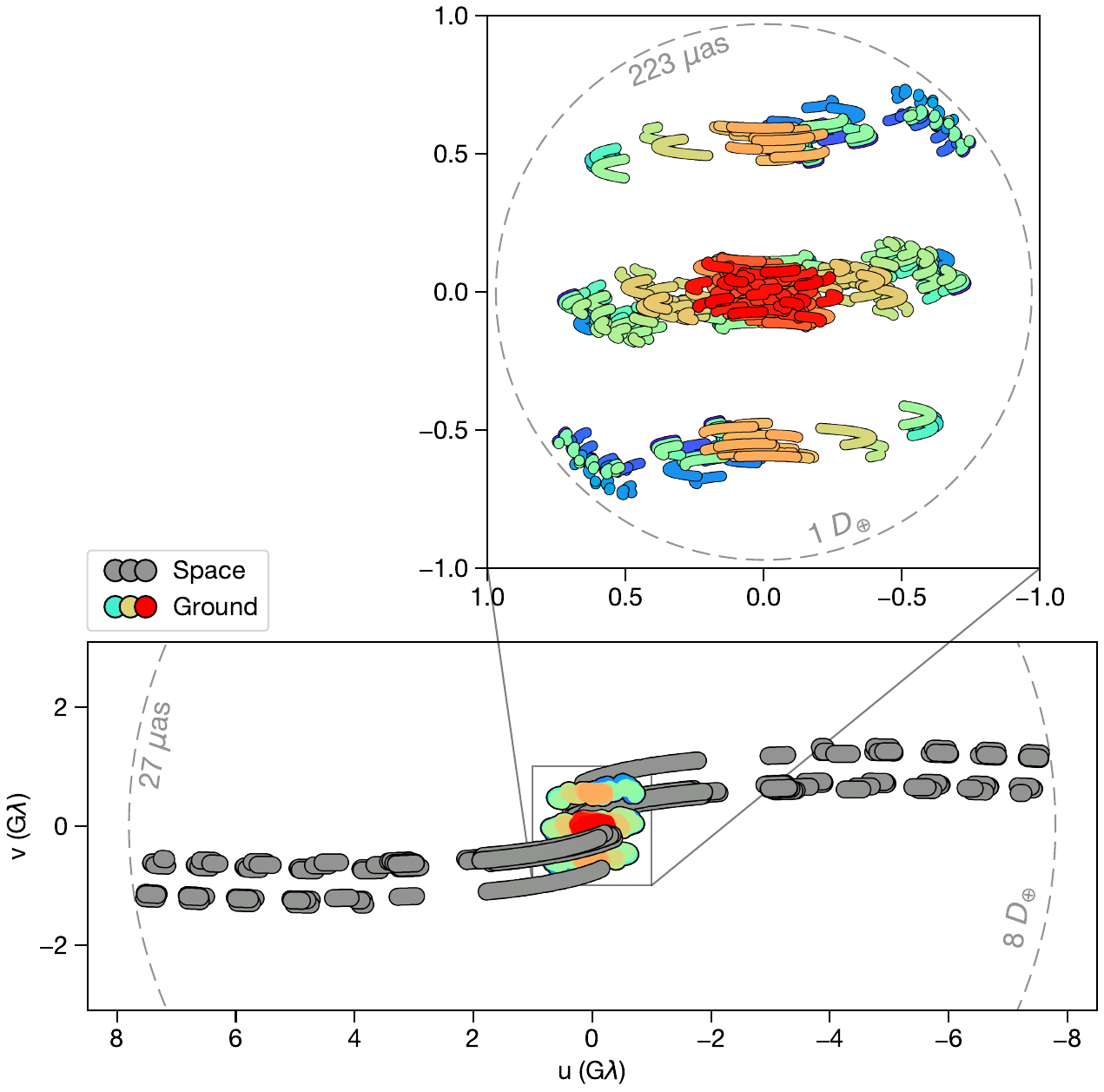}
\caption{{\bf Baseline coverage for our \ra observations of \3c in March 2014.} Rainbow-coloured and grey points indicate individual ground-ground baselines and space-ground baselines, respectively. Dashed circles indicate the baseline length in Earth's diameter units ($D_{\oplus}$) and the corresponding angular resolution.}
\label{fig:uvplot}
\end{extfig*}

\begin{extfig*}
\centering
\includegraphics[width=\textwidth]{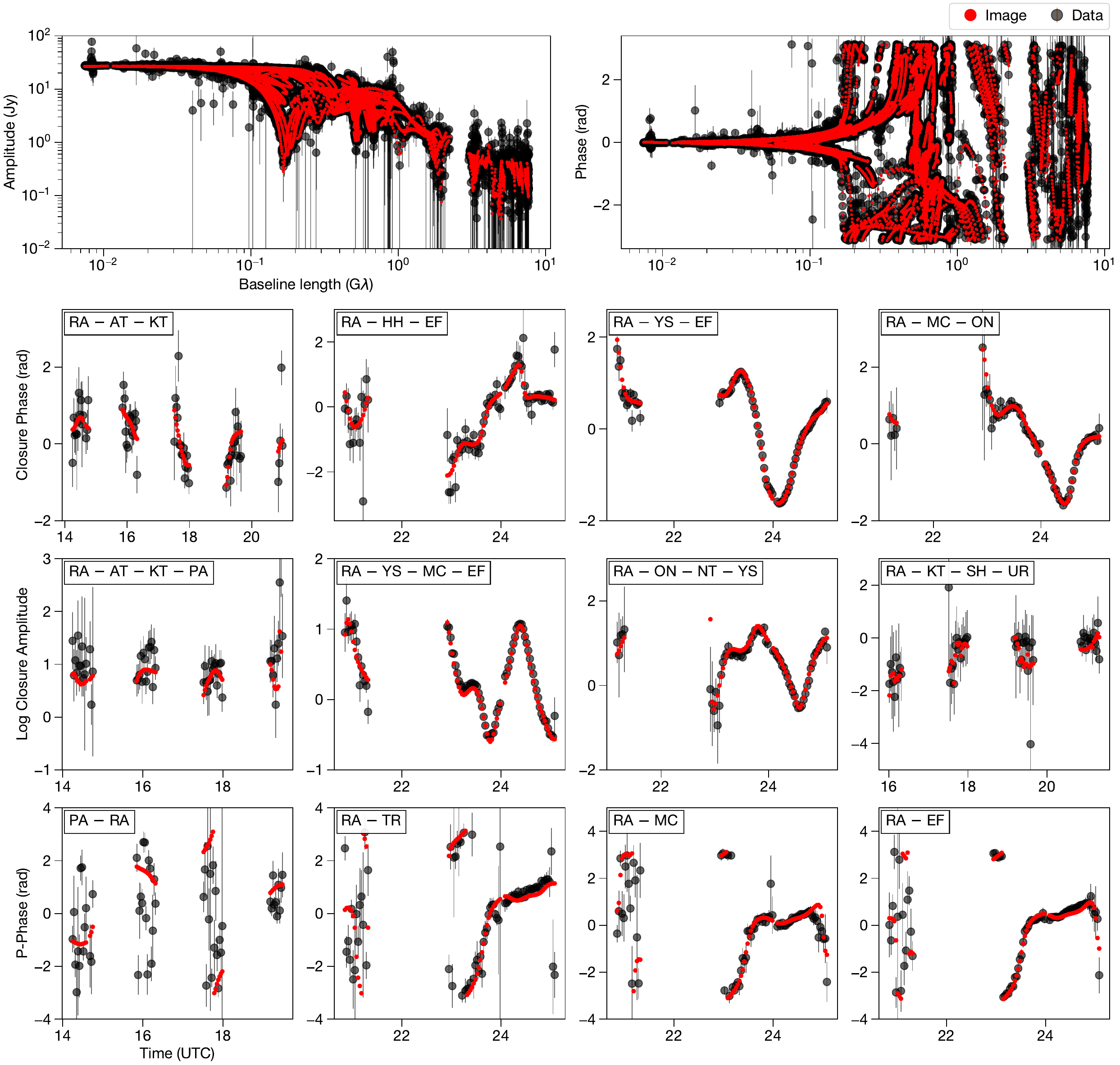}
\caption{{\bf Fitting of the polarimetric \ra image to a selection of data products.} Data (black points) and image model (red points) self-calibrated visibility amplitudes and phases, closure phases, log closure amplitudes, and polarimetric visibility phases as a function of time. All these examples include \ra measurements.}
\label{fig:fitting}
\end{extfig*}

\begin{extfig*}
\centering
\includegraphics[width=\textwidth ]{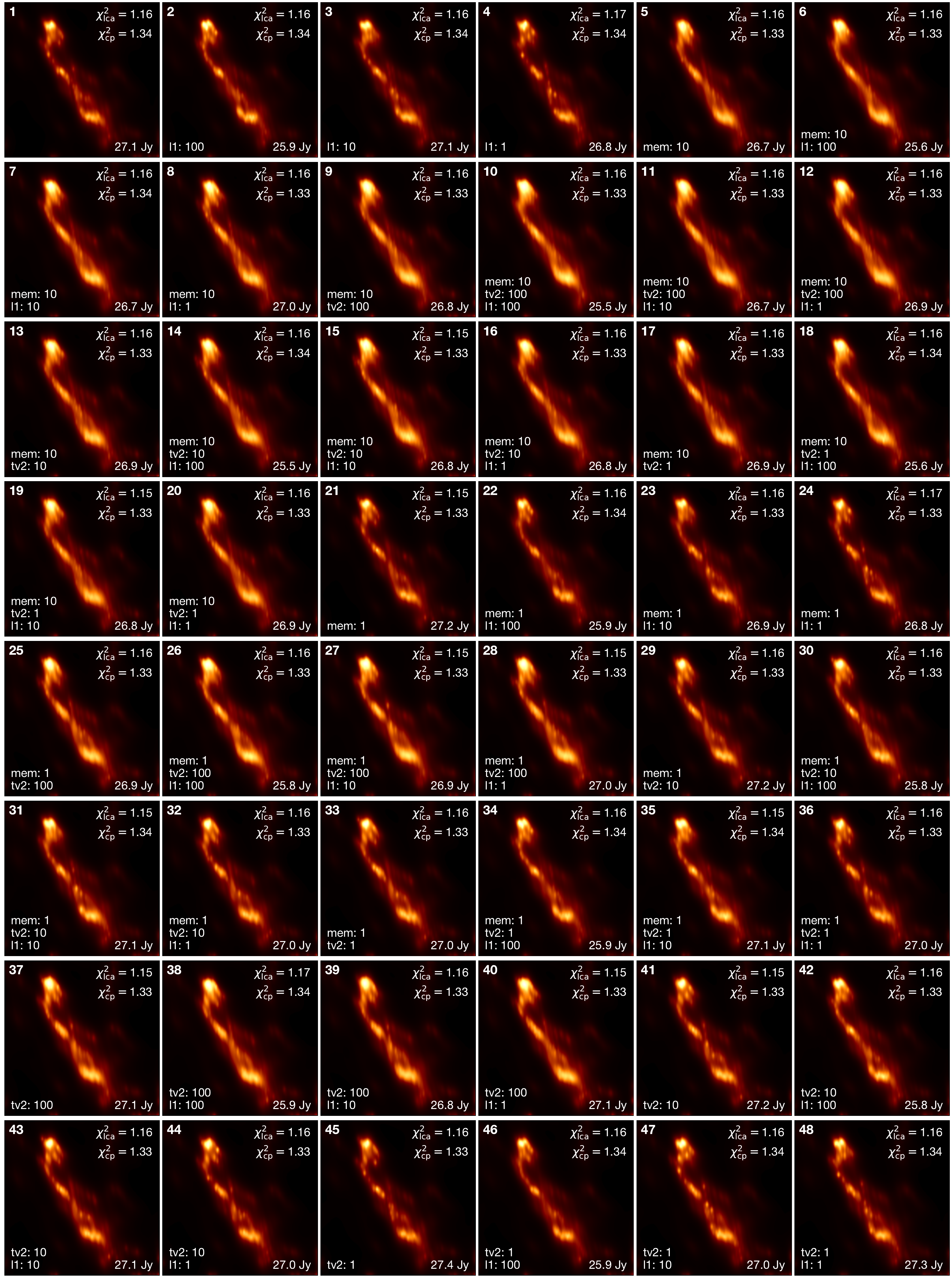}
\caption{{\bf Top 48 image reconstructions from the parameter survey conducted.} Each image includes the closure phase (cp) and log closure amplitude (lca) reduced $\chi^2$, the image regularizers used and their weight, and the total flux reconstructed.}
\label{fig:survey}
\end{extfig*}

\begin{extfig*}
\centering
\includegraphics[width=0.8\textwidth ]{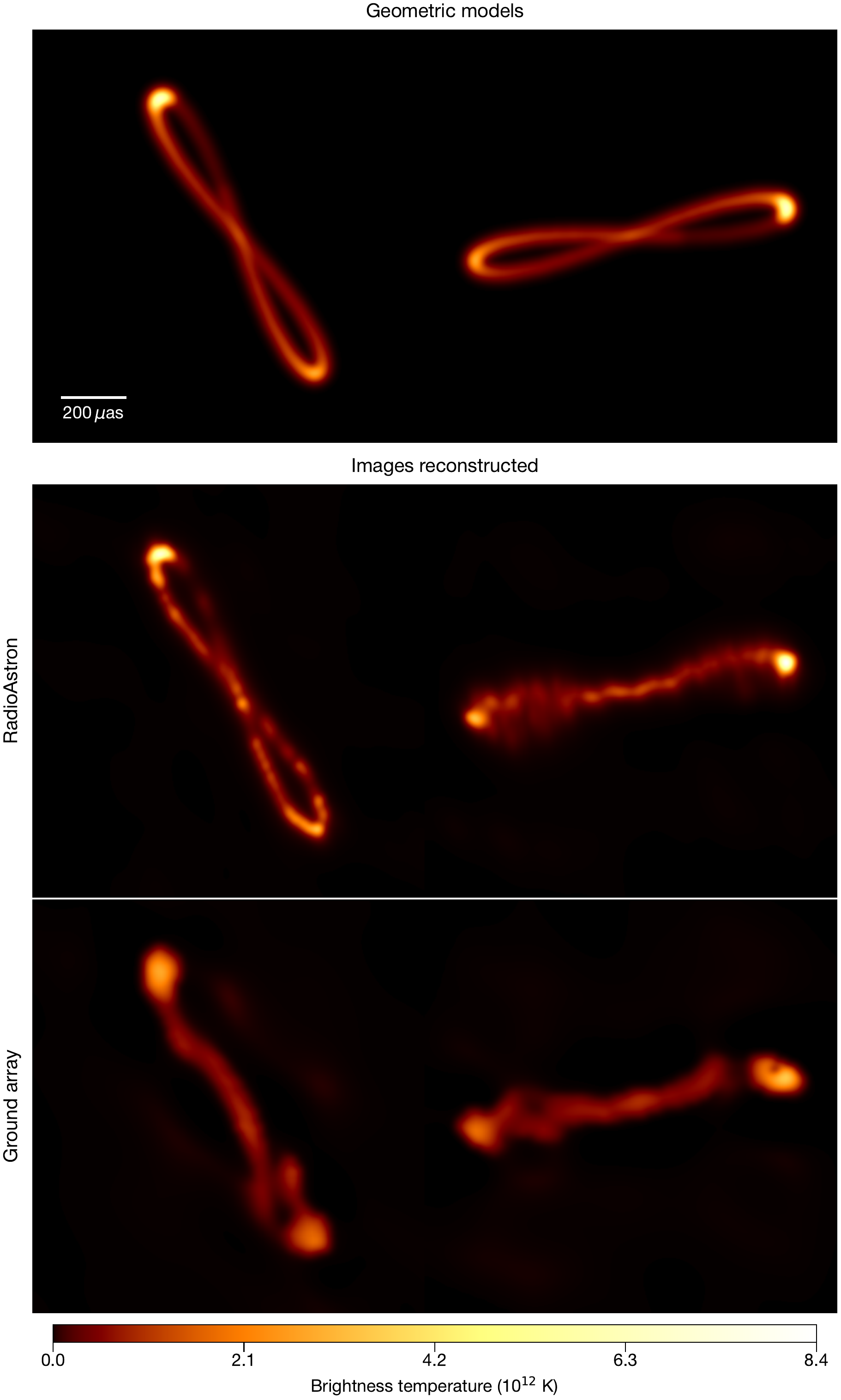}
\caption{Synthetic data tests. In the top row we present the geometric models used to generate the synthetic data. In the middle and bottom rows we show, respectively, the images reconstructed from each data set when \ra is included in the array and when only ground stations participate.}
\label{fig:syn_data}
\end{extfig*}

\end{document}